\documentclass[acmsmall,nonacm]{acmart}

\renewcommand\footnotetextcopyrightpermission[1]{}
\usepackage[normalem]{ulem}

\usepackage{booktabs}   

\usepackage{subfig}

\usepackage{algorithm}
\usepackage[noend]{algpseudocode}
\usepackage{multirow}
\usepackage{multicol}

\begin{document}

\title{Minotaur: A SIMD-Oriented Synthesizing Superoptimizer}

\author{Zhengyang Liu}
\orcid{0000-0003-4938-1935}
\affiliation{%
  \institution{University of Utah}
  \city{Salt Lake City}
  \country{USA}
}
\email{liuz@cs.utah.edu}

\author{Stefan Mada}
\orcid{0009-0006-8473-3667}
\affiliation{%
  \institution{University of Utah}
  \city{Salt Lake City}
  \country{USA}
}
\email{stefan.mada@utah.edu}

\author{John Regehr}
\orcid{0000-0001-7025-4610}
\affiliation{%
  \institution{University of Utah}
  \city{Salt Lake City}
  \country{USA}
}
\email{regehr@cs.utah.edu}

\newcommand{\tool}{Minotaur}

\begin{abstract}
  A superoptimizing compiler---one that performs a meaningful search
  of the program space as part of the optimization process---can find
  optimization opportunities that are missed by even the best existing
  optimizing compilers.
  We created \tool{}: a superoptimizer for LLVM that uses program
  synthesis to improve its code generation, focusing on integer and
  floating-point SIMD code.
  On an Intel Cascade Lake processor, \tool{} achieves an average
  speedup of 7.3\% on the GNU Multiple Precision library (GMP)'s benchmark
  suite, with a maximum speedup of 13\%.
  On SPEC CPU 2017, our superoptimizer produces an average speedup of
  1.5\%, with a maximum speedup of 4.5\% for 638.imagick.
  Every optimization produced by \tool{} has been formally verified,
  and several optimizations that it has discovered have been
  implemented in LLVM as a result of our work.
\end{abstract}

\maketitle

\thispagestyle{empty}

\section{Introduction}

Optimizing compilers emit better code than non-optimizing compilers
do, but even so their output is usually far from optimal.
Our work started when we noticed substantial opportunities for
improvement in the output of LLVM's autovectorizer.
As a step towards fixing these, we created \tool{}: a synthesis-based
superoptimizer for the LLVM intermediate
representation~\cite{LLVM:CGO04} that focuses on LLVM's portable
vector operations as well as its x86-64-specific SIMD intrinsics.
Our goal is to automatically discover useful optimizations that are
missed by LLVM\@.

\tool{} works on code fragments that do not span multiple loop
iterations; it is based on the assumption that existing compiler
optimization passes such as loop unrolling, software pipelining, and
automatic vectorization will create the necessary opportunities for
its optimizations to work effectively.
For example, consider this loop, in C, from the
compression/decompression utility gzip, where \texttt{name} is the
base address of a string and \texttt{p} is a pointer into the string:

{\small\begin{quote}
\begin{verbatim}
do {
  if (*--p == '.') *p = '_';
} while (p != name);
\end{verbatim}
\end{quote}}

When it is compiled by LLVM~18 for a target supporting AVX2
vector extensions, this code is found inside the loop:

{\small\begin{quote}
\begin{verbatim}
%1 = shufflevector <32 x i8> %0, poison, <31, 30, 29, 28, 27, ... 4, 3, 2, 1, 0>
%2 = icmp eq <32 x i8> %1, <46, 46, 46, 46, 46, ... 46, 46, 46, 46, 46>
%3 = shufflevector <32 x i1> %2, poison, <31, 30, 29, 28, 27, ... 4, 3, 2, 1, 0>
\end{verbatim}
\end{quote}}


The first shufflevector reverses a 32-byte chunk of the string, the
\texttt{icmp} instruction checks which elements of the chunk are equal
to 46 (ASCII for the period character), and then the second
shufflevector reverses the vector containing the results of the
computation.
This code cannot be optimized further by LLVM~18; when it is lowered to
object code and executed on an Intel Cascade Lake processor, it
requires 13 uOps, or ``micro-operations,'' processor-internal
RISC-like instructions that modern x86 implementations actually
execute.
\tool{}, on the other hand, automatically determines that the vector
reversals are unnecessary, and rewrites the code in this equivalent,
but significantly cheaper (three uOps), form:

{\small\begin{quote}
\begin{verbatim}
%3 = icmp eq <32 x i8> %0, <46, 46, 46, 46, 46, ... 46, 46, 46, 46, 46>
\end{verbatim}
\end{quote}}


Although SIMD operations are \tool's main focus, it also discovers
optimizations for scalar code.
For example, this code, from the SPEC CPU 2017 benchmark 619.lbm,
computes the difference between two floating-point values, and then
checks if the result is greater than zero:

{\small\begin{quote}\begin{verbatim}
%0 = fsub float %x, %y
%1 = fcmp ogt float %0, 0.000000e+00
\end{verbatim}
\end{quote}}

\tool{} found that this code is equivalent to checking if the second
value is less than the first:

{\small\begin{quote}\begin{verbatim}
%1 = fcmp ogt float %x, %y
\end{verbatim}
\end{quote}}

It is perhaps surprising that LLVM, in 2024, could not perform this
simple rewrite, which reduces the computation cost from seven uOps to
five.
However, it has now been implemented in upstream LLVM as a result of
our work.

\begin{figure*}[tbp]
    \includegraphics[width=\linewidth]{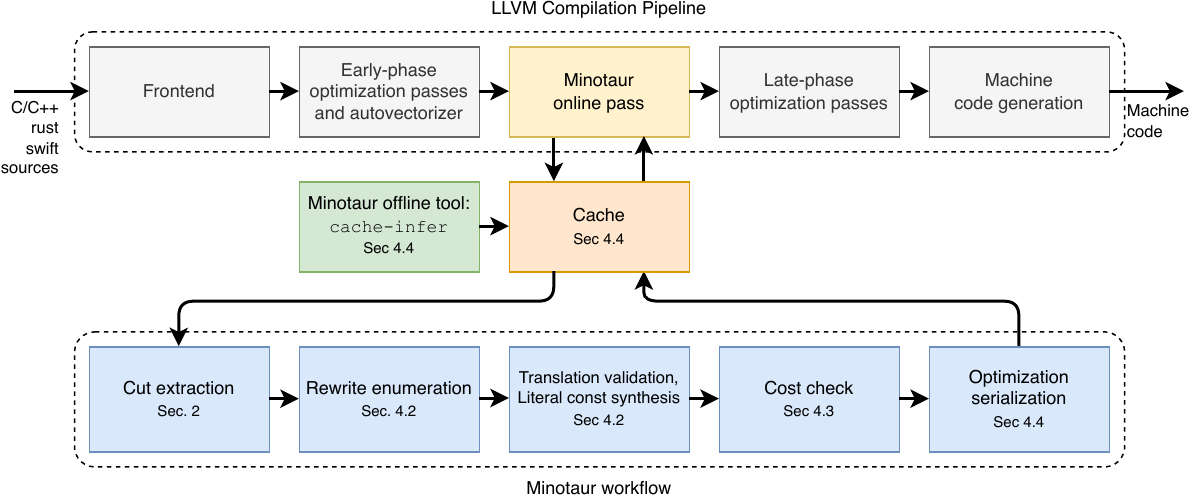}
    \caption{Overview of how \tool{} works, and how it fits into the
      LLVM optimization pipeline}
    \label{fig:workflow}
\end{figure*}

Figure~\ref{fig:workflow} illustrates \tool's high-level structure,
and how it fits into LLVM\@.
It works by extracting many different \textit{cuts} from an LLVM function.
Each cut serves as the specification for a program synthesis
problem, where the objective is to synthesize a new cut that refines
the old one and is cheaper.
When such a cut is found, \tool{} uses it to rewrite the original
LLVM function, and also caches the rewrite.

Reasoning about the correctness of optimizations at the level of LLVM
IR can be very difficult; we have repurposed Alive2~\cite{alive2} to
serve as a verification backend.
To Alive2, we added formal semantics for Intel-architecture-specific
SIMD intrinsics.
Reasoning about the relative costs of code sequences is another
difficult problem; the solution adopted by \tool{} is to reuse the
LLVM Machine Code Analyzer~\cite{llvmmca}, which has adequately
accurate pipeline models for various modern processors.
These tools, along with the LLVM compiler itself, form the
foundation upon which \tool{} is built.

\textbf{Research contributions:}
First, we designed and implemented a domain-specific program
transformer that extracts an SSA value from an LLVM function, along
with context about how that value was computed.
Extracting enough context to permit interesting optimizations, without
extracting so much context that the underlying SMT solver was
overwhelmed, was an interesting empirical problem.
Second, we created a synthesis engine that searches for cheaper code
sequences; it enumerates partially symbolic candidates where the
instructions are concrete, but constants are symbolic.
For this part of our work, we created formal semantics for 165 LLVM
intrinsic functions that correspond to SIMD operations supported by
x86-64 processors, and added these to Alive2.
We also modified Alive2 to support synthesis of literal constants.
Third, to mitigate the large performance overhead of running program
synthesis at compile time, we developed infrastructure for caching
optimizations.
Thus, while \tool{} can be hundreds of times slower than \texttt{clang
  -O3} when its cache is cold, with a warm cache it is just 3\%
slower, when building the SPEC CPU 2017 benchmarks.

We performed a detailed evaluation of \tool's ability to speed up
code, showing that it can find numerous optimizations that LLVM fails
to perform, and also that it can achieve speedups on a variety of
real-world libraries and benchmarks.
\tool{} is also useful for compiler developers, and in fact several
optimizations it has discovered have now been implemented in upstream
LLVM\@.

\section{Cutting LLVM Functions}
\label{sec:cut}

Using a typical function in LLVM IR as the specification, it is not
practical to directly synthesize an optimized version of that
function.
The state of the art in program synthesis simply does not scale up to
the size of LLVM functions found in the wild.
Instead, \tool{} takes a divide-and-conquer approach: we individually
attempt to optimize each instruction in an LLVM function by extracting
a \textit{cut}---a subset of that instruction's dependencies.
If this cut of LLVM instructions can be optimized by program synthesis
then, by the compositionality of refinement, so can the original LLVM
function.
The rest of this section describes this process in more detail.

\subsection{Problem Statement}

Given a function $F$, an instruction $I$ within $F$, and a depth
bound $B$, our goal is to create a new LLVM function $C$ that:
\begin{enumerate}
\item
  is loop-free,
\item
  returns the value computed by $I$, and
\item
  contains every instruction in $F$ that can be reached by following
  up to $B$ backwards data, control, and memory dependency edges.
\end{enumerate}
Informally, we can think of $C$ as summarizing a subcomputation in
$F$, that is (hopefully) tractable for an SMT solver to reason about.

When an instruction is part of $C$ but its inputs are not, they
become free inputs---these are implemented by adding them as
parameters to $C$.
We can think of every instruction that is not part of the cut as being
part of a residual function $R$.
However, note that \tool{} does not explicitly construct $R$---it
computes and optimizes $C$, and then applies the discovered
optimization, if any, directly to $F$ using a rewrite mechanism
described in Section~\ref{sec:rewrite}.

\subsection{Example}

\begin{figure}
  \centering
  \includegraphics[width=0.28\linewidth]{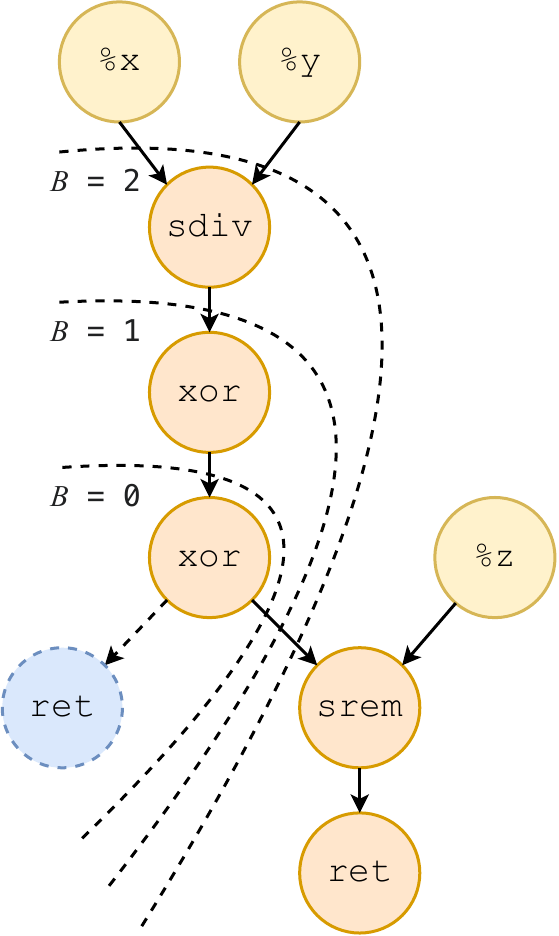}
  \caption{Example of cutting an LLVM function}
  \label{fig:cut-depth}
\end{figure}

Consider this LLVM function that takes three 64-bit arguments and
returns a 64-bit value, where \texttt{sdiv} is signed integer division
and \texttt{srem} is the signed integer modulus operator:


{\small\begin{quote}
\begin{verbatim}
define i64 @f(i64 %x, i64 %y, i64 %z) {
  %a = sdiv i64 %x, %y
  %b = xor i64 %a, -1
  %c = xor i64 %b, -1
  %d = srem i64 %c, %z
  ret i64 %d
}
\end{verbatim}
\end{quote}}

Figure~\ref{fig:cut-depth} illustrates various cuts of this function.
If we cut this function with respect to \texttt{\%c} with $B = 0$ then
we get the following decomposition (however, again, please bear in
mind that \tool{} does not actually construct $R$---we show it here to
make the explanation concrete):

\begin{multicols}{2}
{\small\begin{quote}
\begin{verbatim}
define i64 @r0(i64 %x, i64 %y, i64 %z) {
  %m = sdiv i64 %x, %y
  %n = xor i64 %m, -1
  %o = call i64 %c0(i64 %n)
  %p = srem i64 %o, %z
  ret i64 %p
}
\end{verbatim}
\end{quote}}
\columnbreak
{\small\begin{quote}
\begin{verbatim}
define i64 @c0(i64 %t1) {
  %t2 = xor i64 %t1, -1
  ret i64 %t2
}
\end{verbatim}
\end{quote}}
\end{multicols}

This is not useful, the cut \texttt{c0} contains too little context
to support any optimizations.
If we cut \texttt{f} with respect to \texttt{\%c} with $B = 1$ then we get:

\begin{multicols}{2}
{\small\begin{quote}
\begin{verbatim}
define i64 @r1(i64 %x, i64 %y, i64 %z) {
  %m = sdiv i64 %x, %y
  %o = call i64 @c1(i64 %m)
  %p = srem i64 %o, %z
  ret i64 %p
}
\end{verbatim}
\end{quote}}
\columnbreak
{\small\begin{quote}
\begin{verbatim}
define i64 @c1(i64 %t1) {
  %t2 = xor i64 %t1, -1
  %t3 = xor i64 %t2, -1
  ret i64 %t3
}
\end{verbatim}
\end{quote}}
\end{multicols}

This decomposition is useful: \texttt{c1} can be optimized to simply
return its argument.
If we increase the depth bound to two, then we would get:

\begin{multicols}{2}
{\small\begin{quote}
\begin{verbatim}
define i64 @r2(i64 %x, i64 %y, i64 %z) {
  %o = call i64 @c2(i64 %x, i64 %y)
  %p = srem i64 %o, %z
  ret i64 %p
}
\end{verbatim}
\end{quote}}
\columnbreak
{\small\begin{quote}
\begin{verbatim}
define i64 @c2(i64 %t1, i64 %t2) {
  %t3 = sdiv i64 %t1, %t2
  %t4 = xor i64 %t3, -1
  %t5 = xor i64 %t4, -1
  ret i64 %t5
}
\end{verbatim}
\end{quote}}
\end{multicols}

Here the cut \texttt{c2} can again be optimized (it can just return
\texttt{\%t3}), but now the solver must reason about a 64-bit signed
division---operations like this are difficult and frequently lead to
timeouts.
Choosing an appropriate depth bound is an empirical problem that
we address in Section~\ref{sec:loops}.

\subsection{Correctness Argument}

The composition of $R$ and $C$ is equivalent to the original function:
$F = R \circ C$.
In other words, the decomposition of $F$ into $R$ and $C$ preserves
the behavior of the original function---the difference is simply that
some dependency edges that were previously internal to $F$ now cross
the boundary between $R$ and $C$.

Next, if \tool{} can synthesize $C'$, an optimized function that
refines $C$, then we can compose that with the residual function to
get a new function $F' = R \circ C'$.
Since refinement is compositional, it follows that $F'$ refines $F$,
which is the property that we need for \tool{} to be a correct
optimizer.
The details of establishing a refinement relation between functions in
LLVM IR were presented by Lopes et al.~\citep{alive2}.

Alive2 is intended to be a sound refinement checker for LLVM
IR for LLVM functions that do not contain loops.
We avoid this potential unsoundness by ensuring that $C$ is loop-free,
in which case $C'$ is also loop-free since \tool{} never synthesizes a
loop.

\begin{algorithm}[tbp]
  \small
  \caption{\small{Extract a cut from an LLVM function}}
  \begin{algorithmic}[1]
  \Function{ExtractCut}{{F}: Function, {I}: Instruction, {B}: $\mathbb{N}$}
  \If{$I$ is in a loop}
    \State {{AllowedBasicBlocks} $\gets $ all basic blocks in {I}'s loop (innermost loop if nested)}
  \Else
    \State {{AllowedBasicBlocks} $\gets$ all basic blocks in {F} that is not in a loop}
  \EndIf
  \State {{Harvested} $\gets \emptyset $}
  \State {{Unknown} $\gets \emptyset $}
  \State {{Visited} $\gets \emptyset $}
  \State {{WorkList} $\gets$ \{ ({I}, 0) \}}
  \\
  \\
  \Comment{Phase 1: Instruction extraction}
  \While{{WorkList} is not empty}
  \State {({WI}, {Depth}) $\gets $ {WorkList}.pop()}
  \If {{WI} $\in$ {Visited}}
  \State \textbf{continue}
  \EndIf
  \State {Insert {WI} into {Visited}}
  \If{{Depth} $>$ {B}}
  \State {Insert {WI} into {Unknown}}
  \State {\textbf{continue}}
  \EndIf
  \If{{WI} is not supported}
  \State {Insert {WI} into {Unknown}}
  \State {\textbf{continue}}
  \EndIf
  \State {{BB} $\gets$ {WI}'s basic block}
  \If{{BB} $\notin$ {AllowedBasicBlocks}} 
  \State {Insert {WI} into {Unknown}}
  \State {\textbf{continue}}
  \EndIf
  \State {Insert {WI} into {Harvested}}
  \If{{WI} is a Load instruction}
    \State {{M} $\gets$ {WI}'s linked \texttt{MemoryPhi} or \texttt{MemoryDef}}
    \If {{M} is a \texttt{MemoryDef} $\wedge$ {M} is a store}
    \State {{MI} $\gets$ {M}'s stored value}
    \State {Push ({MI}, {Depth} + 1) into {WorkList} }
  \EndIf

  \Else
    \ForAll{operand {Op} in {WI}}
    \State {Push ({Op}, {Depth} + 1) into {WorkList} }

    \EndFor
  \EndIf

  \State {{T} $ \gets$ terminator of {WI}'s basic block}
    \If {{T} is a conditional branch instruction}
    \State {{TI} $\gets$ {T}'s condition value}
    \State {Push ({TI}, {Depth} + 1) into {WorkList} }
    \EndIf

  \EndWhile
  \State {Insert every terminator instruction in {F} to {Harvested}}
  \\
  \\
  \Comment{Phase 2: Construct a loop-free LLVM function}
  \State {Clone {F} into {C}}
  \State {Delete instructions in {C} except those in {Harvested}}
  \State {Delete all back-edges in {C}}
  \State {Add values in {Unknown} to {C} as function arguments}
  \State {Create return instruction for {I} in {C}}
  \State {\textbf{return} {C}}
  \EndFunction
  \end{algorithmic}
  \label{alg:slicing}
\end{algorithm}

\subsection{Detailed Solution}

Algorithm~\ref{alg:slicing} shows the procedure that \tool{} uses to
extract a cut.
It works in two phases.
In the first, \tool{} determines which instructions will be part
of the cut, using a depth-bounded depth-first search.
During the search, two sets, \emph{Harvest} and \emph{Unknown}, are
propagated which will be used in the second phase for constructing the
cut.
\tool{} uses LLVM's LoopInfo pass~\cite{loopinfo} to identify loops in the
source function.
If instruction $I$ is in a loop, \tool{} will only extract
instructions that are defined inside the loop.
If the loop is nested, \tool{} will only extract instructions that are
defined inside the innermost loop.
\tool{} gives up if the loop is irregular.
If $I$ is not in a loop, \tool{} will skip the instructions that are in a loop.
\tool{} marks non-intrinsic function calls, operations on global
variables, and operations that are not recognized by Alive2 as
unsupported.
All unsupported operations, operations that are beyond the depth
limit, and operations that are outside the loop are discarded and
replaced with free inputs.

For each conditional branch instruction, \tool{} extracts the branch
condition, since these carry control flow information that is useful
during synthesis.
Similarly, when it extracts a load from memory, \tool{} consults
LLVM's MemorySSA pass~\cite{MemorySSA} to get a list of stores that
potentially influence the loaded value.
MemorySSA marks memory operations with one of the three memory access tags:
\texttt{MemoryDef}, \texttt{MemoryUse}, and \texttt{MemoryPhi}.
Each memory operation is associated with a version of the memory state.
A \texttt{MemoryDef} can be a store, a memory fence, or any operation
that creates a new version of the memory state.
A \texttt{MemoryPhi} combines multiple \texttt{MemoryDef}s when
control flow edges merge.
A \texttt{MemoryUse} is a memory instruction that does not modify
memory, it only reads the memory state created by \texttt{MemoryDef}
or \texttt{MemoryPhi}; a load instruction is always a
\texttt{MemoryUse}.
Because it must overapproximate, \tool{} is conservative when finding
load-affecting stores: it starts from the loads
in \texttt{MemoryUse}'s memory version and walks along the MemorySSA's
def-use chain.
When the associated memory operation is a \texttt{MemoryDef}, it
checks if the operation is a store and pushes the stored value into
the worklist, to provide \tool{} a more specific context to optimize
the load instruction.

In the second phase, \tool{} builds the extracted function; it does
this by cloning the original function and then deleting all
instructions that are not in the cut.
\tool{} then deletes all loop backedges, so that the extracted
function is loop-free.
Finally, a return instruction is added to return the value computed by
the instruction that is the basis for the cut.

\begin{figure}
  \centering
  \includegraphics[width=0.8\linewidth]{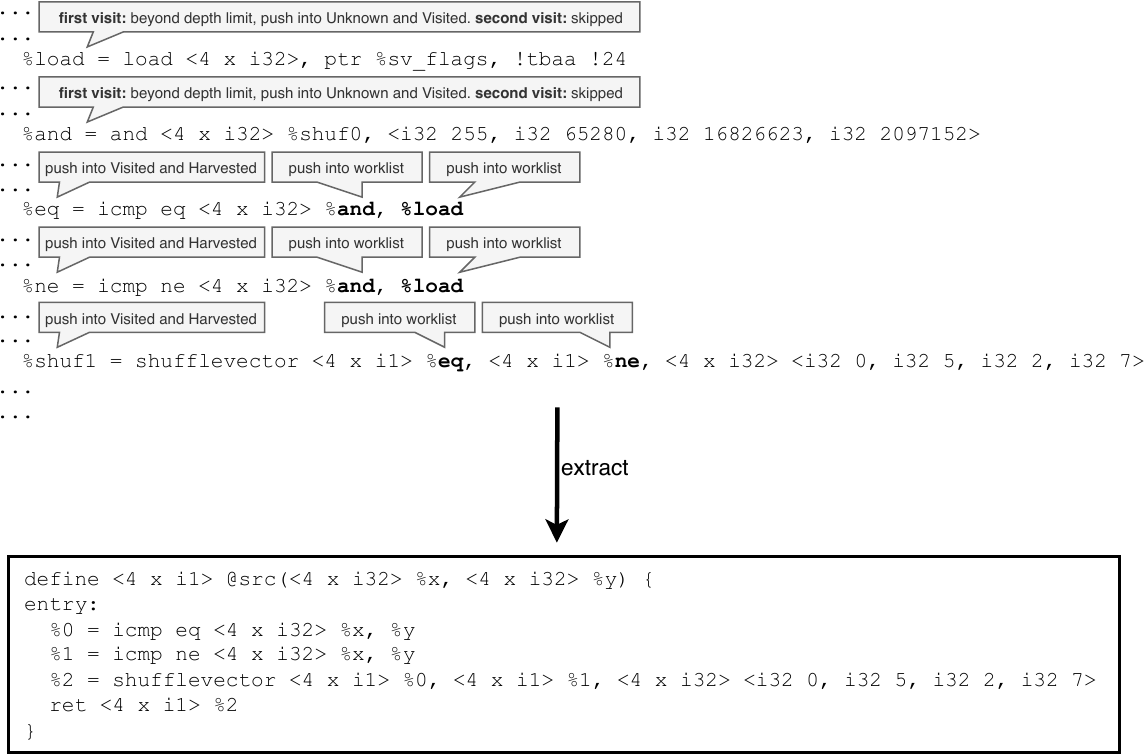}
  \caption{Example of cut extraction}
  \label{fig:cut}
\end{figure}

Figure~\ref{fig:cut} shows an example of cut extraction
for value \texttt{\%shuf1}.
The cutting algorithm starts on \texttt{\%shuf1} and walks along the
def-use chain to extract the instructions that are involved in the
computation of \texttt{\%shuf1}.
A new function is created to hold the extracted instructions shown
in the bottom of the figure.

\subsection{Relation to Previous Cut-Based Superoptimizers}

\tool's cut extraction algorithm is fundamentally more aggressive
than Bansal and Aiken's approach~\cite{Bansal06}, which extracted a
small window of sequential instructions.
It is also considerably more aggressive than Souper~\cite{souper},
which had a very limited view of control flow and refused to consider
memory operations, vector operations, and floating-point operations.

\section{Formalizing Vector Intrinsics in Alive2}

In order for \tool{} to use Alive2 as a verification backend, we had
to modify Alive2 to support a number of x86-64-specific
vector intrinsics.

\subsection{Background: Vectors in LLVM}

LLVM uses a typed, SSA-based intermediate representation (IR).
It supports a derived \emph{vector type}; for example, a vector with
eight lanes, where each element is a 64-bit integer, would have type
\texttt{<8 x i64>}.
Many LLVM instructions, such as arithmetic operations, logical operations,
and pointer arithmetic, can operate on vectors as well as scalars.
IR-level vectors are target-independent; backends attempt to lower
vector operations to native SIMD instructions, if available.

Beyond the vertical ALU instructions that are element-wise vector
versions of scalar instructions, LLVM supports a variety of horizontal
vector reduction intrinsics and an assortment of memory intrinsics
such as vector load and store, strided load and store, and
scatter/gather.
Additionally, there are three vector-specific data movement
instructions:
\textit{extractelement} retrieves the element at a specified index from
a vector;
\textit{insertelement} non-destructively creates a new vector where
one element of an old vector has been replaced with a specified value;
and, \textit{shufflevector} returns a new vector that is a permutation
of two input vectors using elements whose indices are specified by a
constant mask vector.
Finally, to provide direct access to platform-specific vector
instructions, LLVM provides numerous intrinsic functions such as
\texttt{@llvm.x86.avx512.mask.cvttps2dq.512}, aka ``convert with
truncation packed single-precision floating-point values to packed
signed doubleword integer values.''

\renewcommand\algorithmicdo{}
\begin{algorithm}[tbp]
  \small
    \caption{Semantics of \texttt{@llvm.x86.\{avx|avx2|avx512\}.pavg.\{b|w\}}}
    \label{semantic:pavg}
    \begin{algorithmic}[1]
    \State{\textsc{pavg<lanes, bits, masked>}($S_{a}$, $S_{b}$, PassThrough, Mask)}
    \For{each $i$ in range [0 to \textsc{lanes} - 1]}
    \If {\textsc{masked} $\wedge \neg$ Mask[$i$] }
        \State {$S_{ret}$[$i$].val $\gets$ PassThrough[$i$].val}
        \State {$S_{ret}$[$i$].poison $\gets$ PassThrough[$i$].poison}
    \Else
        \State {$S_{ret}$[$i$].val $\gets$ ($S_{a}$[$i$].val $+_\textsc{\tiny{bits}}$ $S_{b}$[$i$].val $+_\textsc{\tiny{bits}}$ 1) /$_\textsc{\tiny{bits}}$ 2}
        \State {$S_{ret}$[$i$].poison $\gets S_{a}$[$i$].poison $\vee$  $S_{b}$[$i$].poison}
    \EndIf
    \EndFor
    \end{algorithmic}
\end{algorithm}

\begin{algorithm}[tbp]
  \small
    \caption{Semantics of \texttt{@llvm.x86.\{sse2|avx2|avx512\}.pmadd.wd}}
    \label{semantic:pmadd}
    \begin{algorithmic}[1]
    \State{\textsc{pmadd.wd<lanes, masked>}($S_{a}$, $S_{b}$, PassThrough, Mask)}
    \For{each $i$ in range [0 to \textsc{lanes}  - 1]}
    \If {\textsc{masked} $\wedge \neg$ Mask[$i$] }
        \State {$S_{ret}$[$i$].val $\gets$ PassThrough[$i$].val}
        \State {$S_{ret}$[$i$].poison $\gets$ PassThrough[$i$].poison}
    \Else
	\State {$S_{ret}$[$i$].val$\gets$sext($S_{a}$[2$\cdot i$].val $\times_\textsc{\tiny{16}}$ $S_{b}$[2$\cdot i$].val) $+_\textsc{\tiny{32}}$  sext($S_{a}$[2$\cdot i+1$].val $\times_\textsc{\tiny{16}}$ $S_{b}$[2$\cdot i +1$].val)}
	\State {$S_{ret}$[$i$].poison$\gets$$S_{a}$[2$\cdot i$].poison $\vee$  $S_{b}$[2$\cdot i$].poison $\vee$ $S_{a}$[2$\cdot i + 1$].poison $\vee$  $S_{b}$[2$\cdot i + 1$].poison}
    \EndIf
    \EndFor
    \end{algorithmic}
\end{algorithm}

\subsection{Assigning a Formal Semantics to Vector Intrinsics}

The version of Alive2 that we started with supports most of the core
LLVM intermediate representation, including its target-independent
vector operations.
However, Alive2 did not have a semantics for any of the numerous
LLVM-level intrinsic functions that provide predictable, low-level
access to target-specific vector instructions.

We added semantics for 165 x86-64 vector intrinsics to Alive2; these
come from the SSE, AVX, AVX2, and AVX-512 ISA extensions.
The resulting version of Alive2 supports the x86 vector intrinsics
that are widely used and that an SMT solver can reason about fairly
efficiently.
This includes special vertical operations that do not overlap with
LLVM's platform-independent vector instructions (such as
\texttt{@llvm.x86.avx2.psign.b}), special data movement intrinsics that
operate differently than LLVM's \texttt{shufflevector} (such as
\texttt{@llvm.x86.avx512.packsswb.512}), and special horizontal
operations that are only available in x86 processors (such as
\texttt{@llvm.x86.avx512.vpdpbusd.512}).
We have not supported dedicated cryptographic operations (that an SMT
solver is unlikely to be able to make use of within a reasonable
amount of CPU time), nor have we supported some unpopular SIMD
intrinsics that we have not observed being used in programs that we
have compiled with \tool{}.

There is significant overlapping functionality between vector
instructions; for example, there are eight different variants of the
\texttt{pavg} instruction that computes a vertical (element-wise)
average of two input vectors.
To exploit this overlap, our implementation is parameterized by vector
width, vector element size, and by the presence of a masking feature
that, when present, uses a bitvector to suppress the output of vector
results in some lanes.
Algorithms~\ref{semantic:pavg} and~\ref{semantic:pmadd} show our
implementations of the \texttt{pavg} (packed average) and
\texttt{pmadd.wd} (packed multiply and add) families of instructions.
This parameterized implementation enabled a high level of code reuse,
and our implementation of these semantics is only 660 lines of C++.
Our semantics differ slightly from the semantics of the corresponding
processor instructions because, at the LLVM level, we must account for
poison values---a form of deferred undefined behavior.
Our strategy for dealing with poison follows the one used by existing
LLVM vector instructions: poison propagates lane-wise, but does not
contaminate non-dependent vector elements.

\subsection{Validating our Changes to Alive2}

We made heavy use of randomized differential testing to ensure that
our new intrinsics correctly implement the intended semantics.
Each iteration of our tester randomly chooses constant inputs to a
single vector intrinsic and then:
\begin{enumerate}
\item
  Creates a small LLVM function passing the chosen inputs to the
  intrinsic.
\item
  Evaluates the effect of the function using LLVM's JIT compilation
  infrastructure~\cite{orc}. The effect is always to produce a
  concrete value, since the inputs are concrete.
\item
  Converts the LLVM function into Alive2 IR and then asks Alive2
  whether this is refined by the output of the JITted code.
\end{enumerate}
Any failure of refinement indicates a bug.
When we fielded this tester, it rapidly found $11$ cases
where our semantics produced an incorrect result, usually for
some edge case.
For example, the semantics for \texttt{pavg} were incorrect when the
sum overflowed.
It also found three cases where \tool{} generated SMT queries that
failed to typecheck.
For example, we set the wrong lane size when parameterizing the
semantics for \texttt{psra.w} and \texttt{psra.d}, causing the solver
to reject our malformed queries.
After we fixed these $14$ bugs, extensive testing failed to find
additional defects.

\section{Synthesizing Optimizations}

For every cut extracted from an LLVM function, \tool{}'s goal is to
synthesize a cheaper way to compute the value returned by that cut.
It does this by enumerating \emph{candidates}---code fragments that
potentially refine the current cut.
When a candidate is found that refines the original cut, \tool{}
consults a cost model.
If the new code is cheaper than the original cut, \tool{} applies the
rewrite to the function that is being optimized.

\subsection{Designing an Appropriate Synthesis Procedure}

A delicate part of designing a practical program synthesis algorithm
is determining how much of the search is pushed to the solver, and how
much searching gets done by code outside the solver.
At one extreme, as the Denali paper~\cite{denali02} points out, we
could simply give the SMT solver a conjecture of the form ``No program
of the target architecture computes P in at most eight cycles.''
If the solver can disprove this conjecture, then its counterexample
will tell us how to compute P in eight or fewer cycles.
This kind of query is asking the solver to do all of the work of
finding a program that disproves the conjecture, including reasoning
about the costs of various alternatives, in a single, complicated
query---this is very heavy lifting.
At the other extreme, we could enumerate completely concrete
candidates, and use the SMT solver only to perform the necessary
refinement checks.
The problem with this approach is literal constants: even a single
64-bit constant in the synthesized code will require us to enumerate
and check $2^{64}$ alternatives; this is clearly infeasible.

We spent a considerable amount of time investigating different points
between these extremes, and finally settled on a design that makes
things as easy as possible for the solver, but without exploring all
possible choices of values for literal constants.
\tool{} creates \emph{partially symbolic} candidates where
instructions are represented concretely, but constants are symbolic.
This gives a reasonably tractable enumeration space without giving up
synthesis power.
Our rationale for this design is that, based on extensive experience
with LLVM and Alive2, a lot of individual refinement checks that we
want to perform---especially those that contain multiplications,
divisions, floating point operations, and pointer indirections---are
already very difficult.

\subsection{Synthesis in Minotaur}

\begin{table*}[tbp]
  \centering
  \small
  \begin{tabular}{ r | l }
    \textbf{Operation Type} & \textbf{Instructions} \\
    \hline
    Unary integer & ctpop, ctlz, cttz, bitreverse, bswap \\
    Unary floating point & fneg, fabs, fceil, ffloor, frint, fround, fnearbyint, froundeven \\
    Binary integer & add, sub, mul, udiv, sdiv, umax, umin, smax, smin\\
    Binary floating point & fadd, fsub, fmul, fdiv, frem, fmaximum, fminimum, fmaxnum, fminnum \\
    Bitwise & and, or, xor, shl, lshr, ashr \\
    Comparison & icmp, fcmp, select \\
    Conversion & zext, sext, trunc, fptrunc, fpext, fptosi, sitofp, fptoui, uitofp \\
    Data movement & extractelement, insertelement, shufflevector \\
    SIMD intrinsics & 165 vector intrinsics mapping to SSE, AVX, AVX2, and AVX-512 instructions \\
  \end{tabular}
  \caption{Operations that \tool{} can synthesize. The data movement
  and SIMD intrinsic instructions require vector operands. The rest
  of the operations apply to both scalar and vector values.}
  \label{tab:operations}
\end{table*}

\begin{algorithm}[tbp]
  \small
  \caption{Minotaur's Synthesis Procedure}
  \label{alg:enumerate}
  \begin{algorithmic}[1]
  \Function{SynthesizeRefinements}{Cut: Function, InstLimit: $\mathbb{N}$, TimeLimit: $\mathbb{N}$}
  \\
  \Comment {Phase 1: Populate the instruction pool}

  \State {Inputs $\gets$ all the SSA definitions in Cut}
  \State{InstPool $\gets \emptyset$}
  \ForAll {op in binary operations listed in Table~\ref{tab:operations}}
    \State {InstPool $\gets$ InstPool $\cup$ \{ (op hole, hole), (op sym-const, hole), (op hole, sym-const) \} }
    \ForAll {input1 in Inputs}
      \State {InstPool $\gets$ InstPool $\cup$ \{ (op input1, sym-const), (op sym-const, input1) \}}
      \State {InstPool $\gets$ InstPool $\cup$ \{ (op input1, hole), (op hole, input1) \}}
      \ForAll {input2 in Inputs}
        \State {InstPool $\gets$ InstPool $\cup$ \{ (op input1, input2) \}}
      \EndFor
    \EndFor
  \EndFor

  \ForAll {other operations in Table~\ref{tab:operations}}
  \Comment{omitted for brevity}
    \State {...}
  \EndFor

  \\
  \\
  \Comment{Phase 2: Generate partially-symbolic candidates}

  \State {WorkList $\gets$ \{ (ret hole), (ret sym-const) \} }

  \State {Candidates $\gets$ Inputs}

  \While {WorkList $\neq \emptyset$}
    \State {I $\gets$ WorkList.pop()}

    \If {I does not contain holes}
      \State {Candidates $\gets$ Candidates $\cup$ \{ I \}}
      \State{\textbf{continue}}
    \EndIf

    \ForAll {Hole in I}
      \If {CountNewInsts(I) $\geq$ InstLimit}
        \State{\textbf{continue}}
      \EndIf
      \ForAll {Inst in InstPool}
        \State {J $\gets$ I with Hole substituted by Inst}
          \If {TargetTransformInfoCost(J) $\geq$ TargetTransformInfoCost(Cut)}
            \State{\textbf{continue}}
          \EndIf

          \State {WorkList $\gets$ WorkList $\cup$ \{ J \}}
      \EndFor
    \EndFor

  \EndWhile

  \\
  \\
  \Comment{Phase 3: Refinement checking and constant synthesis}
  \State {Sort Candidates by TargetTransformInfoCost}
  \State {StartTime $\gets$ time()}
  \State {Refinements $\gets \emptyset$}

  \ForAll {C in Candidates}
    \If {C does not contain symbolic constants}
      \If {Alive2 claims that C refines Cut}
        \State {Refinements $\gets$ Refinements $\cup$ \{ C \}}
      \EndIf
    \Else
      \State {Build exists-forall query to get a model for symbolic constants}
      \If {satisfiable}
        \State {C' $\gets$ C with symbolic constants substituted by the constants in the model}
        \State {Refinements $\gets$ Refinements $\cup$ \{ C' \}}
      \EndIf
    \EndIf
    \If {time() - StartTime $\geq$ TimeLimit}
      \State{\textbf{break}}
    \EndIf
  \EndFor

  \State{\textbf{return} Refinements}

  \EndFunction
  \end{algorithmic}
  \label{alg:synthesis}
\end{algorithm}

Algorithm~\ref{alg:synthesis} describes \tool's synthesis procedure.
In Phase~1, it creates a pool of instructions whose operands are
selected from the available SSA values in the current cut (a dominance
check is not required since cuts are constructed in such a way that
every existing SSA definition dominates the synthesized portion of the
function), from symbolic constants, and from \emph{holes} that
represent instructions that have not yet been enumerated.
The list of instructions that \tool{} can synthesize is shown in
Table~\ref{tab:operations}.
The description in Algorithm~\ref{alg:synthesis} only shows the case
for instructions taking two operands, and it also omits a number of
simple pruning strategies that are useful in practice, such as
avoiding enumeration of redundant versions of commutative operations.
In Phase~2 of the synthesis procedure, instructions from the pool are
used to recursively fill holes; this procedure terminates when all
holes are filled (in which case a complete candidate has been
generated) or when at least one hole remains, but there is no remaining
instruction budget to fill it (in which case the incomplete candidate
is discarded).
A subtlety here is that
LLVM's \texttt{bitcast} instruction, which changes the type
of an SSA value without changing its representation, does not count
towards the instruction limit.
This is because \tool{} takes a low-level, untyped view of values.
For example, it internally treats a 16-way vector of 8-bit values the
same as an 8-way vector of 16-bit values: both of these are simply
128-bit quantities.
This lack of type enforcement allows \tool{} to find interesting,
low-level optimizations such as those that use bitwise operations to
rapidly perform certain floating point operations.

\begin {figure}[tbp]
  \centering
  \includegraphics[width=0.8\linewidth]{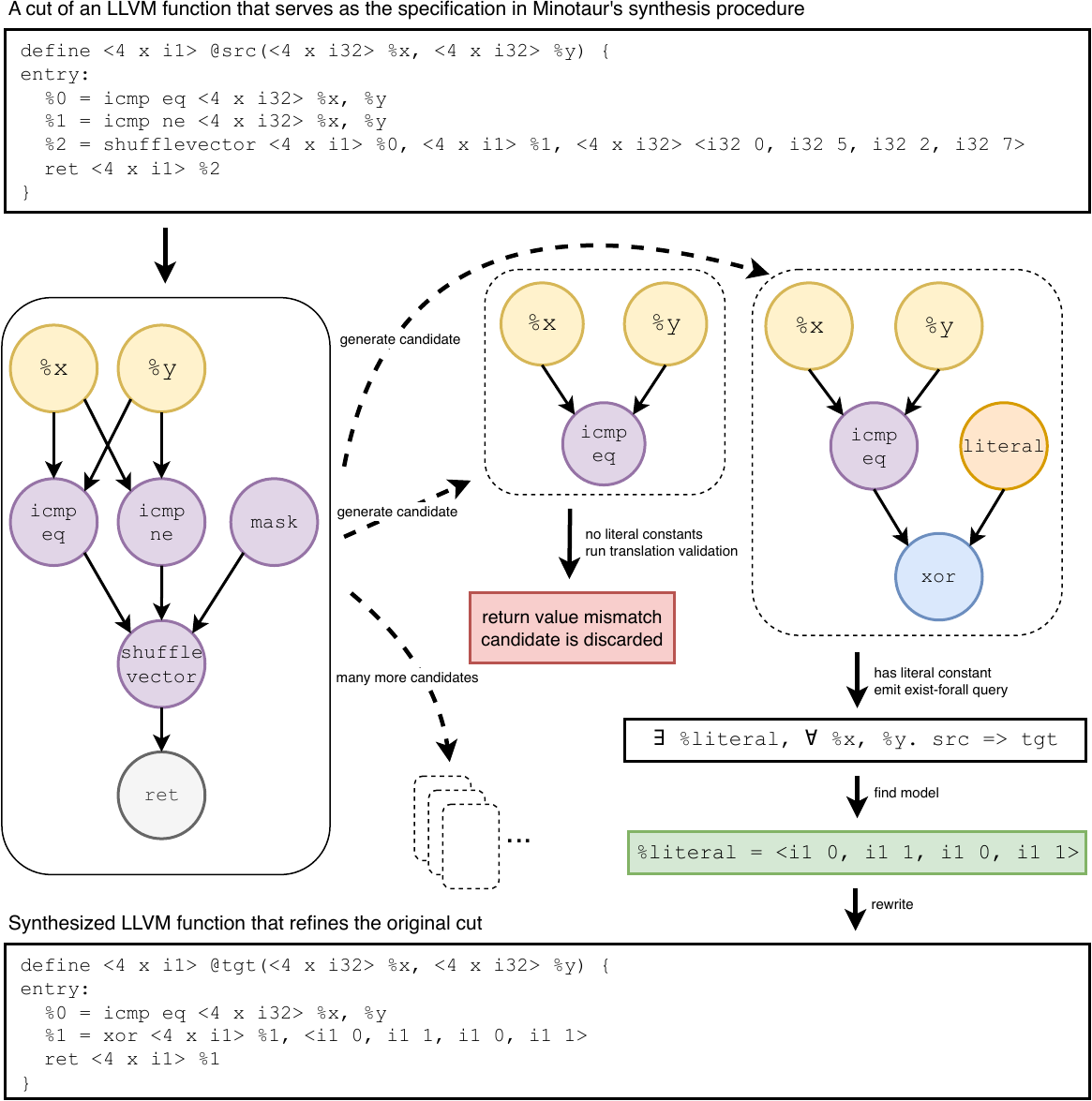}
  \caption{Example of synthesizing a rewrite that contains literal
    constants.  Purple nodes are instructions reused from the original
    cut; blue and orange nodes are synthesized instructions and
    literal constants.}
  \label{fig:synthesizing}
\end{figure}

In Phase~3 of Algorithm~\ref{alg:synthesis}, \tool{} uses Alive2 to
eliminate every candidate that does not refine the specification.
First, we sort the candidates in order of increasing cost using LLVM's
TargetTransformInfo~\cite{tti}: a cost model that roughly captures
execution cost on the target, and is cheap to compute.
We do this to ensure that likely-beneficial rewrites are tested first,
before the synthesis time limit is reached.
For candidates that do not contain symbolic constants, we can use
Alive2 as-is.
To support symbolic constants, we modified Alive2 to wrap
its refinement check in an exists-forall query.
In other words, \tool{} asks the question: ``Does there exist a
valuation of the symbolic constants such that the synthesis candidate
refines the specification for all possible values of the inputs?''
When such a query is satisfiable, the model returned by the solver can
be inspected to find satisfying values of the symbolic constants
in the candidate, which now become literal constants, giving a
complete, sound optimization.
To avoid potentially-expensive exists-forall queries, we experimented
with various techniques such as generalization by
substitution~\cite{Dutertre15}.
However, these failed to outperform exists-forall queries, in the
version of Z3 that we used (4.12.4).
Figure~\ref{fig:synthesizing} illustrates \tool's synthesis procedure.

\subsection{Identifying Profitable Rewrites}

The output of Algorithm~\ref{alg:synthesis} is a list of candidates
that all refine the cut.
Of these, we want to choose the best one---but predicting throughput
of code running on modern microprocessors is not straightforward.
We leverage the LLVM Machine Code Analyzer (LLVM-MCA)~\cite{llvmmca},
which was created to help developers improve performance-critical
code.
It is an interactive tool that emits a graphical depiction of pipeline
behavior, but its functionality can also be accessed programmatically,
and this is what \tool{} does, after lowering each candidate to
x86-64 object code.
Then, \tool{} only applies a rewrite if its estimated cost, using
LLVM-MCA, is lower than that of the original cut, and lower than
that of any other synthesized refinement of the original cut.

Although LLVM-MCA can estimate the cycle cost of LLVM functions, we
instead use the number of uOps (``micro-operations,'' a modern x86-64
processor's internal instruction set) as the estimated cost.
This choice was driven by empirical data: after extensive
experimentation, we determined that, for our purposes, uOps are a
better performance predictor than cycles.

\subsection{Representing and Caching Rewrites}
\label{sec:rewrite}

\tool{} stores each potential rewrite as a pair: $(C, S)$
where $C$ is a cut, represented by a function in LLVM
Intermediate Representation (IR), and $S$ is a rewrite description---an
expression in \tool's own intermediate representation that describes a
different way to compute the return value of $C$.
Rewrite descriptions are directed acyclic graphs containing nodes that represent
operations, and edges representing data flow.
Although the elements found in \tool{} IR are similar to those found
in LLVM IR, we could not reuse LLVM IR to represent rewrites since
LLVM IR does not support incomplete code fragments, and also rewrites
must contain enough information to support connecting the new code in
the rewrite to code in the unoptimized function.

To support caching, rewrites must be serializable.
The cut $C$ can be serialized using existing LLVM functionality, and we
created a simple S-expression syntax for serializing the $S$ part.
Figure~\ref{fig:syntax} shows the syntax of the IR\@.
For example, if the returning value of $C$, a 32-bit instruction is
replaced by left shift by one bit position, the textual format for
the expression is \texttt{(shl (val i32 \%0), (const i32 1), i32)}.

Rewrites are cached in a Redis instance: this implementation choice
allows the cache to be persistent across multiple \tool{} runs and
also makes the cache network-accessible.
Synthesis can be done online---during compilation---but also
offline, in a mode where \tool{} extracts cuts into the Redis
cache but does not perform synthesis.
In this mode, compilation is only slowed down by a few percent.
\tool's offline mode is designed for batch processing.
In this mode, a separate program called \texttt{cache-infer} retrieves
cuts from the cache, runs synthesis on them, and stores any
optimizations that it discovers back into the cache.
Unlike the online mode, which runs synthesis tasks one after the
other, offline mode can run all synthesis jobs in parallel.

\begin{figure}[tbp]
  \small
  \begin{tabular}{r c l}
    \emph{Op} &::=& \emph{Inst} | \emph{Constant} | \emph{Value} \\
    \emph{Inst}  &::=& (\emph{UnaryOp} \emph{Op}, \emph{Type}) | (\emph{BinaryOp} \emph{Op}, \emph{Op}, \emph{Type}) | (\emph{Conversion} \emph{Op}, \emph{Type}) |\\
              && (insertelement \emph{Op}, \emph{Op}, \emph{Op}) | (extractelement \emph{Op}, \emph{Op}) | (shufflevector \emph{Op}, \emph{Op}, \emph{Constant}) |\\
              && (\emph{Comparison} \emph{Op} \emph{Op}) | (select \emph{Op}, \emph{Op}, \emph{Op}) | (\emph{Intrinsic} \emph{Op}, \emph{Op}) \\
    \emph{Constant} &::=& (const \emph{Type} \texttt{number-literal}) \\
    \emph{Value} &::=& (val \emph{Type} \texttt{llvm-identifier}) \\

    \emph{Type} &::=& \emph{ScalarType} | <elements $\times$ \emph{ScalarType}> \\
    \emph{ScalarType} &::=& i1 | i8 | i16 | i32 | i64 | half | float | double | fp128 \\
    \emph{BinaryOp} &::=& xor | and | or | add | sub | mul | udiv | sdiv | ashr | lshr | shl | umax | umin | smax | smin\\
                && fadd | fsub | fmul | fdiv | copysign | fmaximum | fminimum | fmaxnum | fminnum \\
    \emph{UnaryOp} &::=& ctpop | ctlz | cttz | bswap | bitreverse | ret\\
                      && fneg | fabs | fceil | ffloor | frint | fround | ftrunc | fnearbyint | froundeven \\
    \emph{Conversion} &::=& zext | sext | trunc |\\
                    && fptrunc | fpext | fptosi | sitofp | fptoui | uitofp \\
    \emph{Comparison} &::=& eq | ne | ult | ule | slt | sle |\\
                && oeq | ogt | oge | olt | ole | one | ord | ueq | ugt | uge | ult | ule | une | uno \\
    \emph{Intrinsic} &::=& ssse3.phadd.d.128 | avx2.pavg.b | avx512.pmaddubs.w.512 | \dots (165 intrinsics in total) \\
  \end{tabular}
  \caption{Syntax for Minotaur rewrites}
  \label{fig:syntax}
  \Description[syntax]{Minotaur syntax}
\end{figure}

\subsection{Integration with LLVM}

\tool{} is loaded into LLVM as a shared library where it runs as an
optimization pass.
We arranged for it to run at the end of LLVM's auto-vectorization pipeline.
We invoke LLVM's Dead Code Elimination pass after Minotaur to
clean up the resulting code.

\section{Evaluation}
\label{sec:evaluation}

Our primary evaluation metric for \tool{} is its ability to speed up
legacy application code, compared to an optimized build using LLVM~18.
Secondarily, we look at \tool's impact on compile time, optimizations
that have been integrated into upstream LLVM based on our work, and
other issues.

\subsection{Correctness}

Every optimization discovered by \tool{} has been formally verified by
Alive2.
Even so, bugs might remain in the instruction semantics that we have
added to Alive2, in our cut extractor, in our rewrite mechanism, in
Alive2, or in Z3\@.
To defend against implementation errors, we have compiled numerous
open source applications using \tool, and then run those applications'
test suites, to ensure that they were not miscompiled.
Furthermore, we have compiled SPEC CPU 2017 using \tool{} and
used the SPEC drivers to ensure that all of its benchmarks behave
as expected.

\subsection{Effect of Depth Bounds in the Cut Extractor}
\label{sec:loops}




It is important for \tool{} to extract cuts that are of an appropriate
size.
If they are too large, compile times suffer and also the SMT solver
can be overwhelmed, leading to timeouts; if cuts are too small, then
they form an insufficient basis for driving an optimization.
To determine a good value for $B$, the depth parameter to the cut
extraction procedure shown in Algorithm~\ref{alg:slicing}, we
performed an empirical study.
We started with FlexC's benchmark suite~\cite{woodruff2023rewriting},
a collection of 2,386 compilable, non-trivial C functions containing
loops from FFMPEG, FreeImage, DarkNet, xz, bzip2, and the LivermoreC
benchmark.
%
When compiled to LLVM IR, these functions contain a total of 123,062
instructions; thus, our cut extractor was invoked 123,062 times for
each depth bound.
We chose this code as the basis for our experiment because it is
derived from real applications while also being small enough to
keep compile times manageable (compared to, e.g., SPEC CPU 2017,
which is much larger).

\begin{figure*}[tbp]
  \centering
  \subfloat[Unique cuts extracted\label{fig:loop-expression}]{
    \includegraphics[width=0.32\linewidth]{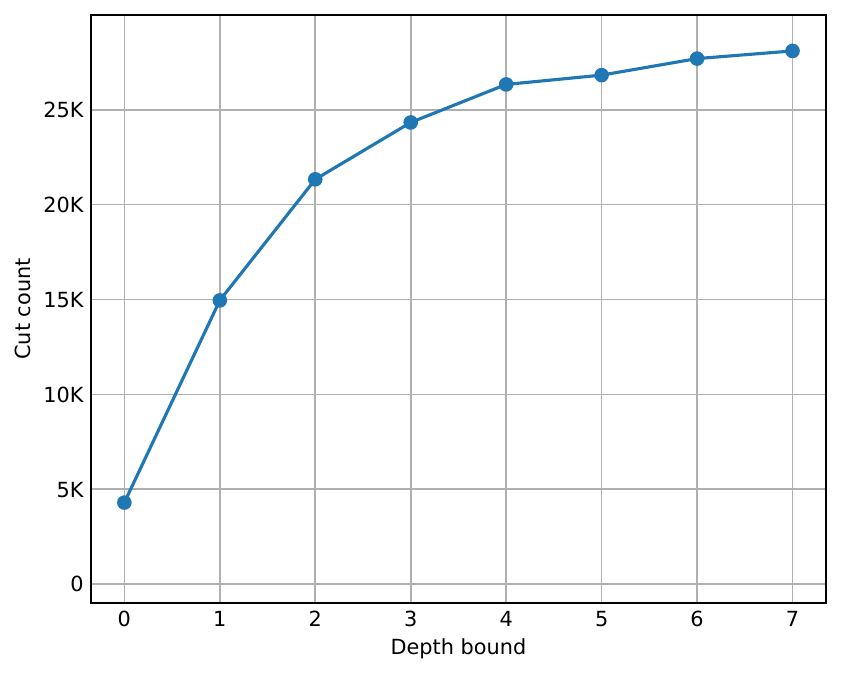}
  }
  \hfill
  \subfloat[Unique opts. synthesized\label{fig:loop-optimization}]{
    \includegraphics[width=0.32\linewidth]{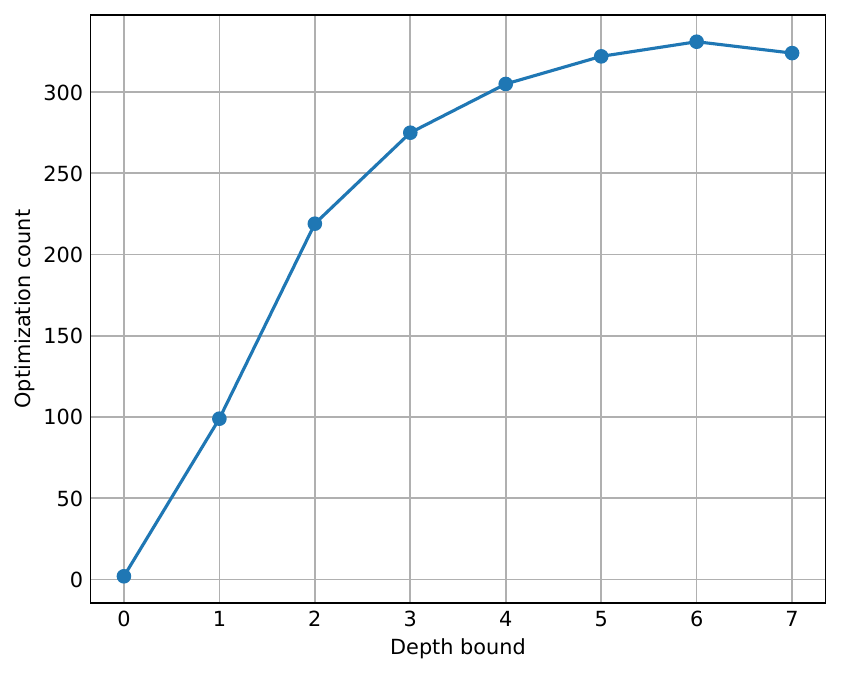}
  }
  \hfill
  \subfloat[Compilation time\label{fig:loop-buildtime}]{
    \includegraphics[width=0.32\linewidth]{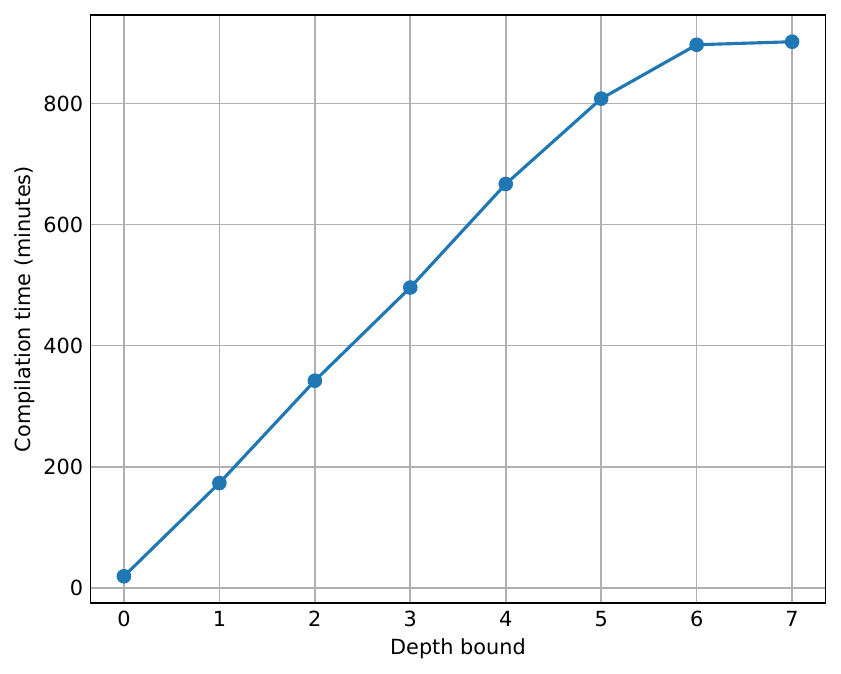}
  }
  \caption{Evaluating the effect of varying $B$, the depth bound for
    cut extraction}
  \label{fig:loop}
\end{figure*}

We then ran \tool{} on these functions with all depth bounds from
0--7, measuring the number of unique cuts that were extracted, the
number of optimizations found, and the compilation time.
We used a one-minute timeout for individual Z3 queries, and we also
gave \tool{} a total of up to five minutes to synthesize an optimized
version of each cut.
Figure~\ref{fig:loop} summarizes the results of this experiment.
The number of unique cuts that are extracted grows quickly with $B$,
but eventually begins to saturate simply because the functions being
compiled do not always have very long dependency chains.
The number of synthesized optimizations also grows quickly, but it
peaks when $B=6$ and then it decreases because the size of the cuts
causes many solver timeouts.
Finally, the total compile time increases smoothly with the depth
bound, eventually leveling off as most solver queries time out.

For the experiments in the rest of the evaluation section, we chose
$B=4$ because this gets close to the maximum observed number of
optimizations without requiring exorbitant compile times.
It seems likely that there is room for improvement in this aspect of
\tool: perhaps the depth bound should be determined adaptively.
In this scenario, we would extract more and more components into the
cut, until either an optimization is found or else the solver begins
to time out.
We leave explorations of this nature for future work.

\subsection{What Kind of Optimizations Matter Most?}

\begin{table*}[t]
  \centering
  \Small
  \begin{tabular}{ r | r r r r r r}
    & memory & FP vector & FP scalar& integer vector & integer scalar & overall \\
    \hline
    Number of rewrites & 3 & 17 & 4 & 191 & 109 & 324 \\
    Geomean speedup & 1.0605x & 1.0600x & 1.0572x & 1.0142x & 1.0506x & 1.0610x \\
    Contribution to speedup & 0.65\% & 1.64\% & 5.90\% & 75.57\% & 16.23\% & 100\% \\
  \end{tabular}
  \caption{Results of an ablation study based on optimization categories}
  \label{tab:ablation}
\end{table*}

To determine which of \tool's optimizations matter most, we performed
an ablation study, again using the FlexC benchmark suite that we
described in Section~\ref{sec:loops}.
We split the optimizations that \tool{} found into five
categories: memory, floating-point vector, floating-point scalar,
integer vector, and integer scalar.
Then, we ran \tool{} in a way that omitted each of these categories of
optimizations.
As shown in Table~\ref{tab:ablation}, integer vector optimizations
produce the most rewrites, and also produce the majority of the
observed speedup.

\subsection{Speedups for Benchmarks and Applications}

In this section, we show how \tool{} speeds up real-world benchmarks
and applications.

\paragraph{Experimental setup}
We used two machines for our evaluation.
The first has an Intel Xeon Gold 6210U processor running at 2.5\,GHz,
and has 20 cores; this implements the Cascade Lake
microarchitecture~\cite{cascadelake} and supports the AVX-512
instruction set.
The second has an AMD Ryzen 5950X processor running at 3.4\,GHz, and
has 16 cores; this processor implements the Zen~3
microarchitecture~\cite{zen3}.
Both machines run Linux and were idle except for a single core running
our benchmarks (however, when measuring compile times, as reported in
Table~\ref{tab:compiletime}, we used all cores).
To reduce the performance variation caused by frequency scaling, we
disabled turbo boost on the Intel machine and core performance
boost on the AMD machine.
We also disabled simultaneous multithreading on both machines.

We invoked LLVM with the \texttt{-march=native} compilation flag to
ask it to take maximum advantage of processor features; we left other
compilation flags unchanged, except where noted.
All benchmarks are compiled at the \texttt{-O3} optimization level.
We set the timeout for Z3~\cite{z3} queries to one minute.
Finally, for each instruction that it tries to optimize, \tool{} gives
up if no solution is found within five minutes.

\paragraph{Benchmark selection}
We evaluate on SPEC CPU 2017\footnote{\url{https://www.spec.org/cpu2017/}}
because it is a widely accepted standard
benchmark.
We only evaluate on the \emph{speed} subset of the SPEC suite, and we omit
648.exchange, 607.cactuBSSN, 621.wrf, 627.cam4, 628.pop2, 649.fotonik3d,
and 654.roms as they contain Fortran code.
We additionally use GMP, the GNU Multiple Precision
Library,\footnote{\url{https://gmplib.org/}} and
libYUV,\footnote{\url{https://chromium.googlesource.com/libyuv/libyuv/}}
which is used by Google Chrome/Chromium for manipulating images in the
YUV format.
We chose these libraries because they have been heavily tuned for
performance, they are loop-intensive, and they come with performance
benchmark suites that we could simply reuse.

\paragraph{Compile times}
Table~\ref{tab:compiletime} shows how long it takes \tool{} to build
our benchmarks, along with the number of potentially optimizable
values and the number of optimizations found.
The compile times are for parallel builds; we set the \textsc{make}'s
\texttt{-j} flag and SPEC CPU 2017's \texttt{build\_ncpus}
configuration to the number of cores on the machine.
Minotaur is very slow when it runs with a cold cache because it
performs many solver queries.
However, with a warm cache, it is only 3\% slower than baseline \texttt{clang}.

In most cases, \tool{} found more optimizations when targeting the AMD
processor.
We believe this is because LLVM is more mature targeting
AVX2 than AVX-512.
Queries with 256-bit vectors are also less likely to timeout in Z3 than
are queries with 512-bit vectors.

\begin{table*}[t]
  \centering
  \Small
  \begin{tabular}{r | r r r | r r | r r r | r r}
    \multirow{2}{*}{} & \multicolumn{5}{c|}{Intel Cascade Lake} & \multicolumn{5}{c}{AMD Zen 3} \\
    & \multicolumn{3}{c|}{Compile time (minutes)} & \multicolumn{2}{c|}{Stats} & \multicolumn{3}{c|}{Compile time (minutes)} & \multicolumn{2}{c}{Stats}  \\
    Benchmarks & Cold & Warm & Clang & \# Cuts & \# Opts. & Cold & Warm & Clang & \# Cuts & \# Opts. \\
    \hline
    SPEC CPU 2017 & 2,337 & 3 & 3 & 109,177 & 2,683 & 2,580 & 3 & 3 & 114,612 & 2,820 \\
    gmp-6.2.1 & 440 & < 1 & < 1 & 9,170 & 336 & 445 & < 1 & < 1 & 9,265 & 387\\
    libYUV & 2,196 & < 1 & < 1 & 6,849 & 334  & 2,193 & < 1 & < 1 & 6,809 & 357 \\
  \end{tabular}
  \caption{\tool's effect on compilation time}
  \label{tab:compiletime}
\end{table*}

\paragraph{Optimizing GMP with \tool{}}

\begin{figure*}[tbp]
  \centering
  \subfloat[Speedups on Intel Cascade Lake, geomean = 1.073x\label{plot:gmp-intel}]{
    \includegraphics[width=\linewidth]{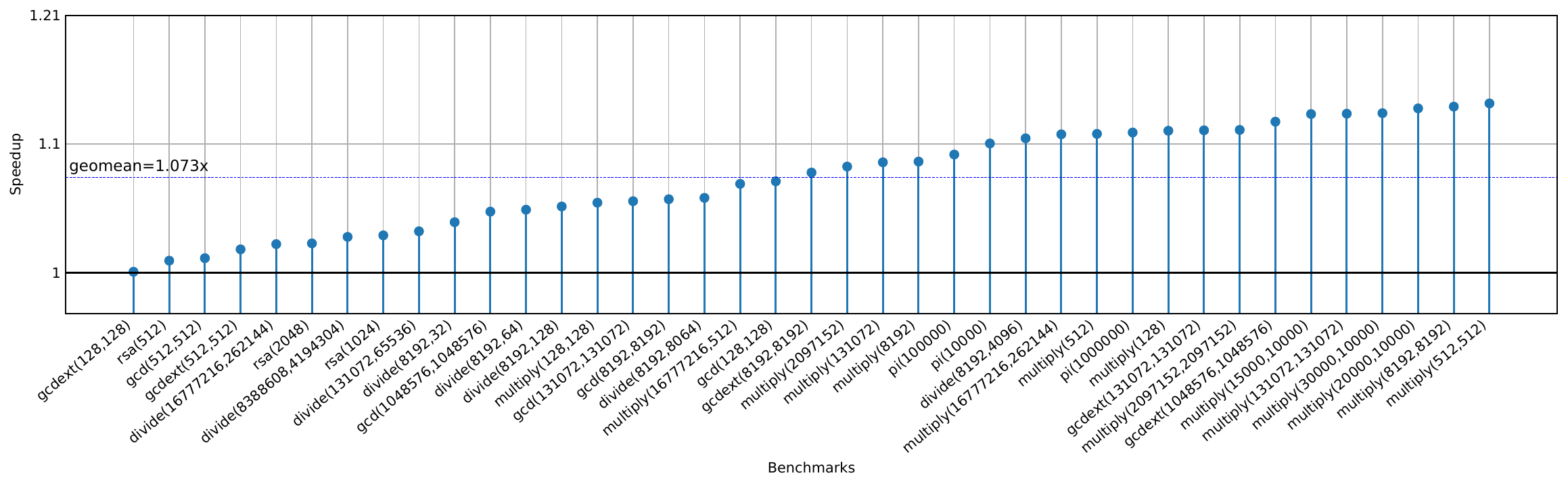}
  }
  \hfill
  \subfloat[Speedups on AMD Zen 3, geomean = 1.065x\label{plot:gmp-amd}]{
    \includegraphics[width=\linewidth]{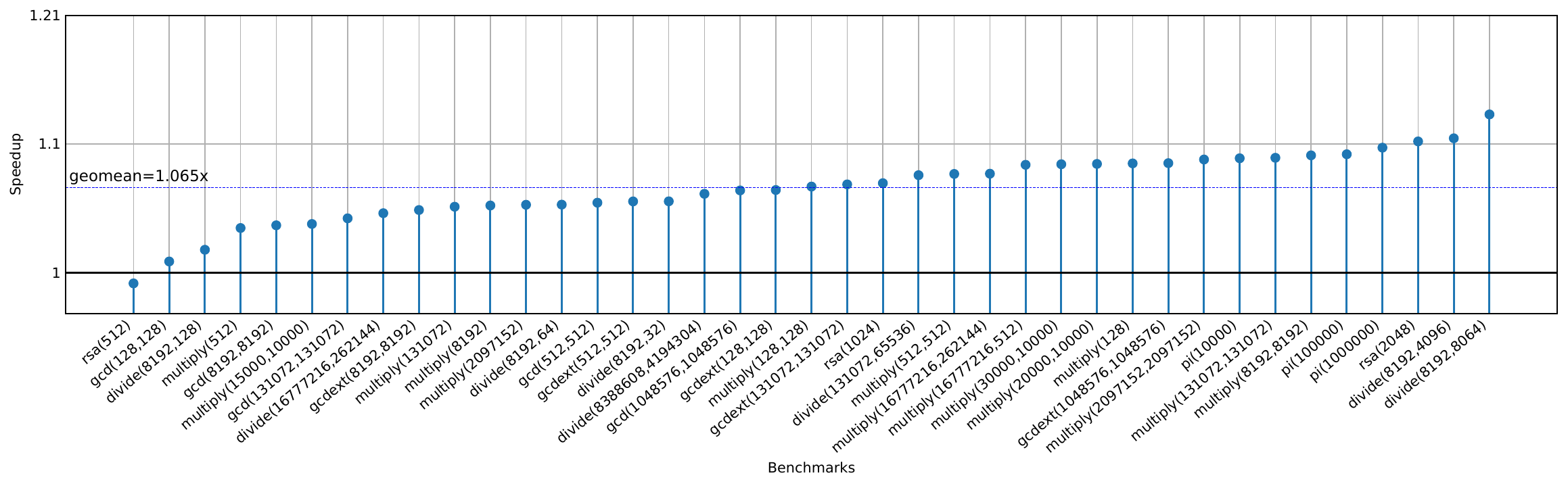}
  }
  \caption{GNU Multiple Precision Library (GMP) speedups, on a logarithmic scale}
  \label{fig:gmp}
\end{figure*}

GMP provides a portable C-language implementation and then, for
several platforms, a faster assembly language implementation.
For this evaluation, we selected the C implementation, because \tool{}
works on LLVM IR and cannot process assembly code at all.
The benchmark suite that we used is
GMPbench.\footnote{\url{https://gmplib.org/gmpbench}}
Figure~\ref{fig:gmp} summarizes the results.
When \tool{} targets the Intel Cascade Lake processor, and when the
resulting executables are run on that same microarchitecture,
all the benchmarks sped up;
across all of the benchmarks, the mean speedup was 7.3\%.
The analogous experiment using the AMD Zen~3 microarchitecture
resulted in one benchmark slowing down, and the rest of benchmarks
speeding up, for an overall mean speedup of 6.5\%.

\paragraph{Optimizing libYUV with \tool{}}

\begin{figure*}[tbp]
  \centering
  \subfloat[Speedups on Intel Cascade Lake, geomean = 1.022x\label{plot:libyuv-intel}]{
    \includegraphics[width=\linewidth]{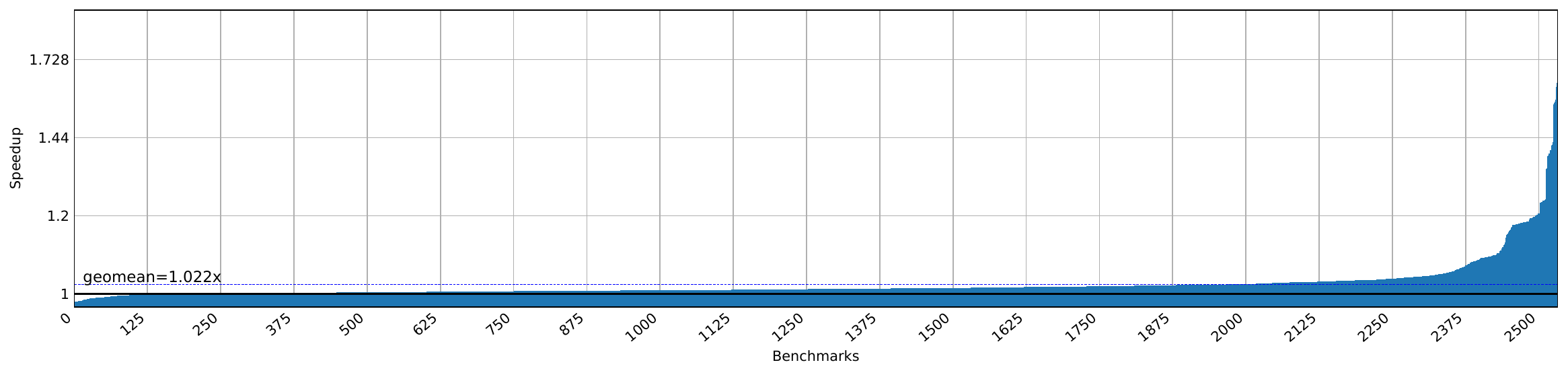}
  }
  \hfill
  \subfloat[Speedups on AMD Zen 3, geomean = 1.029x\label{plot:libyuv-amd}]{
    \includegraphics[width=\linewidth]{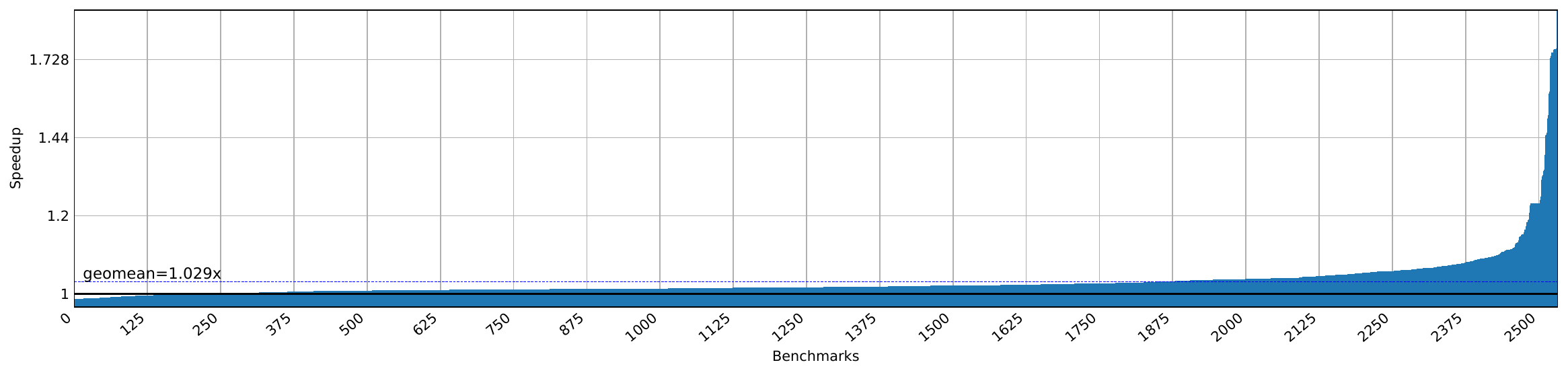}
  }
  \caption{LibYUV speedups, on a logarithmic scale}
  \label{fig:yuv}
\end{figure*}

This library has an extensive test suite, part of which is explicitly
intended for performance testing; we used this part as a benchmark.
Each test program scales, rotates, or converts a 1280\,x\,728 pixel
image 1,000 times.
Figure~\ref{fig:yuv} shows the results of this experiment.
When \tool{} targets an Intel processor, $148$ programs slowed down, $72$
did not change performance, and $2,312$ sped up, for an overall speedup of
2.2\%.
Targeting an AMD processor, $188$ programs slowed down, $85$ did not
change performance, and $2,259$ sped up, for an overall speedup of 2.9\%.
\tool{} can make code slower because it looks at optimizations in
isolation; it does not attempt to model interactions between
optimizations.

libYUV is portable code, but it has already been heavily tuned for
performance; most commits to its repository over the last several
years have been performance-related.
Our hypothesis is that this manual tuning has already eaten up most of
the performance gains that we would have hoped to gain from \tool{}.
For some time now, Google's released versions of Chrome have been
compiled using LLVM; the Chrome engineers have had ample time to
ensure that this compiler achieves decent code generation for
performance-critical libraries.


\begin{figure*}[tbp]
  \centering
  \subfloat[Speedups on Cascade Lake\label{fig:spec-intel-speed-ups}]{
    \includegraphics[width=0.48\linewidth]{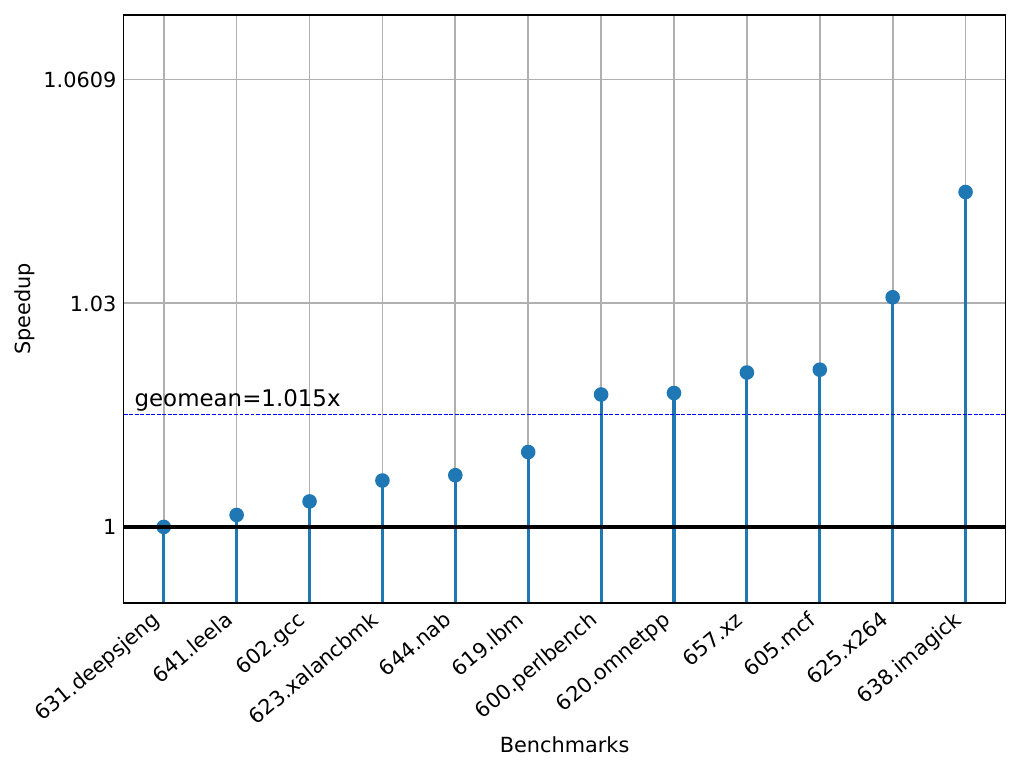}
  }
  \hfill
  \subfloat[Speedups on Zen 3\label{fig:spec-amd-speed-ups}]{
    \includegraphics[width=0.48\linewidth]{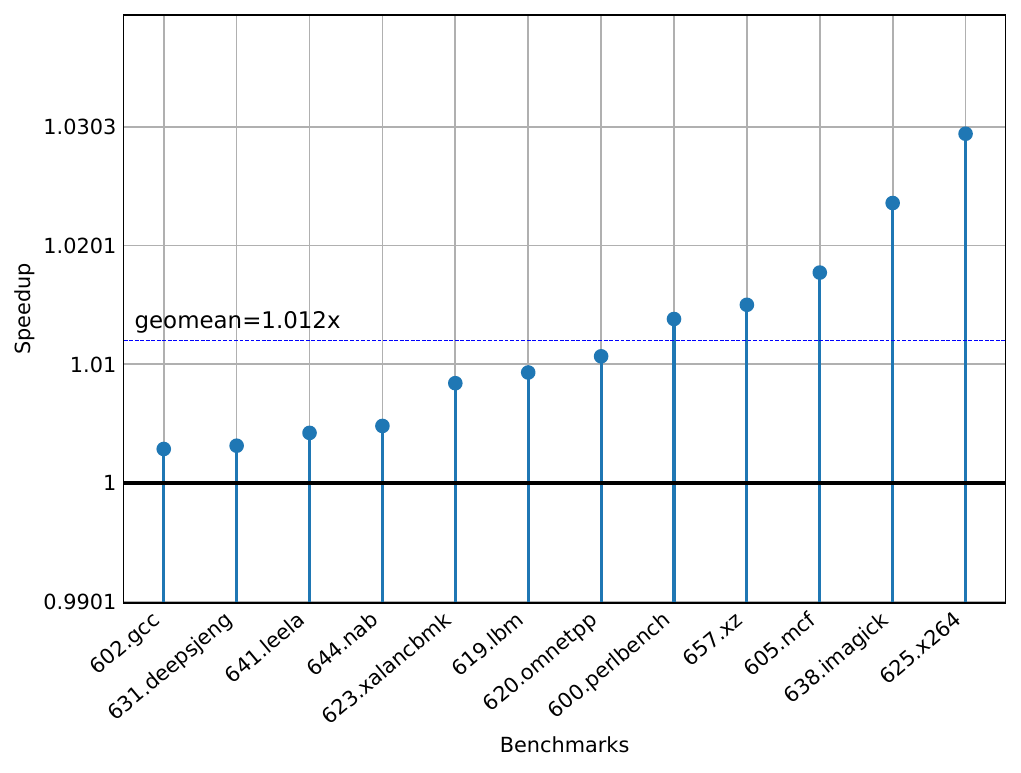}
  }
  \caption{SPEC CPU2017 benchmark performance, on a logarithmic scale}
  \label{fig:spec}
\end{figure*}

\paragraph{Optimizing SPEC CPU 2017 with \tool{}}

Figure~\ref{fig:spec} shows the effect of optimizing the benchmarks
from SPEC CPU2017 using \tool.
When optimizing for, and running on, the Intel processor, we observed
a mean speedup of 1.5\%.
When optimizing for, and running on, the AMD processor, we observed a
mean speedup of 1.2\%.
It is notoriously difficult to speed up the SPEC CPU benchmarks
because compiler engineers have already put considerable effort into
achieving good code generation for them.

\subsection{Impact on Upstream LLVM}

In several cases where an optimization discovered by \tool{} seemed to
be simple and broadly applicable, we have reported its absence as an
LLVM defect, using the project's issue tracker.
This section summarizes the results of this informal LLVM-improvement
project.

We reported ten missing floating-point optimizations.
Five of these (including the one we presented in Section~1) have now
been implemented in LLVM\@.
Three of them are in the code review phase: a patch exists and is
being discussed by developers.
Finally, two of them are being discussed, but a candidate patch does
not yet exist.

We also reported five missing vector optimizations.
One of these has been fixed, one has a patch that is under review, and
three are still being discussed.

\subsection{Optimizations Discovered by \tool}
\label{sec:examples}

The purpose of this section is to examine \tool's strengths by
presenting some optimizations that it found while compiling benchmark
programs.
None of these optimizations can be performed by the version of LLVM
that \tool{} is based on,\footnote{\tool{} uses LLVM~18.1.0 for all
results in this paper.}  at its \texttt{-O3} optimization level.
We present optimizations in an SSA format that is close to LLVM IR,
but we have edited it slightly for compactness and legibility.

\paragraph*{Example 1}

This code is from perlbench in SPEC:

{\small\begin{quote}\begin{verbatim}
%0 = zext <16 x i8> %x to <16 x i16>
%1 = zext <16 x i8> %y to <16 x i16>
%2 = call @llvm.x86.avx2.pavg.w(%0, %1)
%3 = trunc <16 x i16> %2 to <16 x i8>
ret <16 x i8> %3
  =>
%0 = call @llvm.x86.sse2.pavg.b(%x, %y)
ret <16 x i8> %0
\end{verbatim}
\end{quote}}

The unoptimized code zero-extends each 8-bit element of the two input
vectors to 16~bits, calls the AVX2 variant of \texttt{pavg} to perform
element-wise averaging of the extended vectors, and then truncates
elements of the resulting vector back to eight bits.
The optimized code simply calls an SSE2 version of the \texttt{pavg}
instruction that operates on 8-bit elements, reducing the uOp cost
of the operation from four to one.

\paragraph*{Example 2}

This code is from libYUV:

{\small\begin{quote}\begin{verbatim}
%0 = call @llvm.x86.avx2.pmadd.wd(%x, <0,1,0,1, ...>)
%1 = call @llvm.x86.avx2.pmadd.wd(%x, <1,0,1,0, ...>)
%2 = sub nsw <8 x i32> %1, %0
ret <8 x i32> %2
  =>
%0 = call @llvm.x86.avx2.pmadd.wd(%x,<1,-1,1,-1, ...>)
ret <8 x i32> %0
\end{verbatim}
\end{quote}}

The \texttt{pmadd.wd} (multiply and add packed integers) instruction multiplies
signed 16-bit integers element-wise from two input vectors, and then
computes its output by adding adjacent pairs of elements from the
resulting vector.
Thus, the input to this instruction is two 16-way vectors containing
16-bit elements, and its output is a single 8-way vector of 32-bit
elements.

In this example, the second argument to each \texttt{pmadd.wd}
instruction in the unoptimized code is a vector of alternating zeroes
and ones, which has the effect of selecting odd-indexed elements into
\texttt{\%0} and even-indexed elements into \texttt{\%1}.
Then, after the \texttt{sub} instruction, which simply performs
element-wise subtraction of \texttt{\%0} and \texttt{\%1}, the overall
effect of this code is to compute the difference between adjacent
pairs of elements of \texttt{\%x}.
\tool{} is able to perform this same computation using a single
\texttt{pmadd.wd} instruction which negates odd-numbered elements of
\texttt{\%x} before performing the addition.
The optimized code requires $5$ uOps to execute whereas the original
code requires $8$.

\paragraph*{Example 3}

This code is from libYUV:

{\small\begin{quote}\begin{verbatim}
%0 = shufflevector <32 x i8> %x, poison, <3, 7, 11, 15, 19, 23, 27, 31>
%1 = lshr %0, <6, 6, 6, 6, 6, 6, 6, 6>
%2 = zext 8 x i8> %1 to <8 x i32>
ret <8 x i32> %2
  =>
%0 = bitcast <32 x i8> %x to <8 x i32>
%1 = call @llvm.x86.avx2.psrli.d(<8 x i32> %0, 30)
ret <8 x i32> %1
\end{verbatim}
\end{quote}}

The \texttt{shufflevector} instruction in the unoptimized code selects
every fourth byte-sized element from the input \texttt{\%x}.
The resulting 8-way vector is right-shifted element-wise by six bit
positions, and that result is zero-extended to an 8-way vector of
32-bit elements.
\tool's optimized version (which executes in 4 uOps instead of 11)
first reinterprets the input vector's data as 32-bit elements; this
bitcast is relevant to LLVM's type system, but it is a nop at the CPU
level.
Then, the \texttt{prsli} instruction shifts each 32-bit element to the
right by 30 bit positions.
This right-shift-by-30 achieves the same effect as the unoptimized
code, where the \texttt{shufflevector} can be seen as a
right-shift-by-24, followed by an explicit right-shift-by-6.

\paragraph*{Example 4}

This code, from compiling perlbench from SPEC CPU 2017, illustrates
\tool's ability to reason about control flow:

{\small\begin{quote}\begin{verbatim}
entry:
  br i1 %c, label %body, label %if.end
body:
  br label %if.end
if.end:
  %p1 = phi [ %a, %body ], [ %b, %entry ]
  %p2 = phi [ %b, %body ], [ %a, %entry ]
  %r = call @llvm.x86.avx2.pavg.b(%p1, %p2)
  ret <32 x i8> %r
    =>
  %r = call @llvm.x86.avx2.pavg.b(%a, %b)
  ret <32 x i8> %r
\end{verbatim}
\end{quote}}

The intent of the code is to compute the element-wise average of input
vectors \texttt{\%a} and \texttt{\%b}, with a Boolean value
\texttt{\%c} determining the order in which the input vectors are
presented to the \texttt{pavg} instruction.
However, the order of arguments to this instruction does not matter, and
\tool's version executes in 4 uOps while the original code requires
10.
Note that \tool{} was not explicitly taught that \texttt{pavg} is
commutative; the necessary information was inferred naturally from the
formal specification.

\paragraph*{Example 5}

This is an optimization discovered
by \tool{} when it was used to compile GMP:


{\small\begin{quote}\begin{verbatim}
%0 = lshr i64 %x, 1
%1 = and i64 %0, 0x5555555555555555
%2 = sub i64 %x, %1
%3 = lshr i64 %2, 2
%4 = and i64 %2, 0x3333333333333333
%5 = and i64 %3, 0x3333333333333333
%6 = add nuw nsw i64 %4, %3
%7 = lshr i64 %6, 4
%8 = add nuw nsw i64 %7, %6
%9 = and i64 %8, 0xf0f0f0f0f0f0f0f
ret i64 %9
  =>
%0 = bitcast i64 %x to <8 x i8>
%1 = call @llvm.ctpop(<8 x i8> %0)
%2 = bitcast <8 x i8> %1 to i64
ret i64 %2
\end{verbatim}
\end{quote}}

%
%
The original code performs a series of bit-level
manipulations on a 64-bit integer value, with the net result of
performing an 8-way vectorized 8-bit popcount operation.\footnote{The
popcount, or Hamming weight, of a bitvector is the number of ``1''
bits in it.}
The optimized code simply calls an intrinsic function to do the
popcount; it costs 13 uOps instead of the original code's 19.
Although robust recognition of open-coded idioms is not the focus
of our work, \tool{} does sometimes manage to achieve this.

Taking a strict view of types in the synthesis process could help
prune the search space, but it would also cause us to miss
optimizations that require a flexible view of types.
This example illustrates the latter case: the original code contains
no indication that a good optimization can be found using a vector of
type <8 x i8>, and therefore a strictly type-guided synthesis
procedure would miss this one.

\paragraph*{Example 6}

This code comes from 644.nab in SPEC CPU 2017:

{\small\begin{quote}\begin{verbatim}
%0 = fcmp oge float %x, 0.000000e+00
%1 = fneg float %x
%2 = select i1 %0, float %0, float %2
%3 = fcmp oeq float %2, 0.000000e+00
ret i1 %3
  =>
%1 = fcmp oeq float %x, 0.000000e+00
ret i1 %oeq
\end{verbatim}
\end{quote}}

The original code computes the absolute value of a floating-point
number \texttt{\%x} and then checks if the result is zero.
\tool{} found that that the original code is equivalent to simply checking if
\texttt{\%x} is zero.

\paragraph*{Example 7}

This code comes from 619.lbm in SPEC CPU 2017:


{\small\begin{quote}\begin{verbatim}
%0 = fmul float %x, 0x3FF0CCCCC0000000
%1 = fcmp olt float %t1, 0x3FE20418A0000000
ret i1 %1
  =>
%0 = fcmp ole float %x, 0x3FE12878E0000000
ret i1 %0
\end{verbatim}
\end{quote}}

The original code multiplies a floating-point value \texttt{\%x} by a
constant, and then checks if the result is less than another constant.
\tool{} found that this code is equivalent to checking if \texttt{\%x}
is less than or equal to a third constant.
This example shows that \tool{} can reason about and synthesize floating
point literals.

\paragraph*{Example 8}

This code comes from 638.imagick in SPEC CPU 2017:

{\small\begin{quote}\begin{verbatim}
%0 = fmul float %x, 0.000000e+00
%1 = fmul float %0, 3.000000e+00
ret float %1
  =>
%0 = fmul float %x, 0.000000e+00
ret i1 %0
\end{verbatim}
\end{quote}}

The original code multiplies a floating-point value \texttt{\%x} by
zero, and then multiplies the result by 3.0. \tool{} found that this
code is equivalent to multiplying \texttt{\%x} by zero directly.
Note the original code cannot be optimized to 0.0 directly, because of
the NaN and signed zero propagation rules in floating-point arithmetic.
This example shows that \tool{} is able to reason about these corner
cases and synthesize the correct code.

\paragraph*{Example 9}

This code comes from FlexC's benchmark suite:

{\small\begin{quote}\begin{verbatim}
%0 = extractelement <4 x ptr> %0, i32 0
%1 = extractelement <4 x ptr> %0, i32 3
%2 = load i32, ptr %0
%3 = load i32, ptr %1
%4 = insertelement <4 x i32> zeroinitializer, i32 %2, i32 0
%5 = insertelement <4 x i32> %4, i32 %3, i32 3
ret <4 x i32> %5
  =>
%0 = call @llvm.masked.gather(%0, 4, <true, false, false, true>, zeroinitializer)
ret <4 x i32> %0
\end{verbatim}
\end{quote}}

The original code extracts two pointers from a vector of pointers,
loads the values from these pointers, and then inserts these values
into a vector of integers. \tool{} found that this code is equivalent
to performing a masked gather operation, which loads values from
memory using a vector of pointers and a mask.

\section{Related Work}

A \emph{superoptimizer} is a program optimizer that meaningfully
relies on search to generate better code, in contrast with traditional
compilers that attempt a fixed (but perhaps very large) sequence of
transformations.
The eponymous superoptimizer~\cite{massalin} exhaustively generated
machine instruction sequences, using various strategies to prune the
search space, and using testing to weed out infeasible candidates.
Also predating modern solver-based methods, Davidson and Fraser~\cite{peep84}
constructed peephole optimizations from machine description files.
In contrast, modern superoptimizers rely on solvers to perform
automated reasoning about program semantics.

Souper~\cite{souper} is a synthesizing superoptimizer that works on
LLVM IR; it is the most directly connected previous work to \tool{}.
Souper's slicing strategy is similar to \tool's in that it extracts a
DAG of LLVM instructions that overapproximates how a given SSA value
is computed.
However, unlike Souper, \tool{} extracts memory operations and
multiple basic blocks, so it is capable of (we believe) strictly more
transformations than Souper is able to perform.
Additionally, Souper's undefined behavior model does not capture all
of the subtleties of undefined behavior in LLVM, whereas we reuse
Alive2's model, which is the most widely used formalization of these
semantics, and the one that is most widely recognized as being
correct.
Finally, \tool{} focuses on vector-related transformations, whereas
Souper supports neither LLVM's portable vector instruction set nor its
platform-specific intrinsics.
It is worth noting that, over the years, the LLVM developers have
implemented numerous optimizations discovered by Souper.
These are all, of course, present in LLVM~18, the compiler that is the
baseline for our experimental evaluation.
In other words, \tool{} is an effective superoptimizer on top of a
previous solver-based superoptimizer (and Souper was effective on top
of an even earlier LLVM superoptimizer~\cite{Sands11}).

\tool{} is also strongly inspired by Bansal and Aiken's
work~\cite{Bansal06}; their superoptimizer operated on x86 assembly
code and was able to make interesting use of vector instructions.
Starting from unoptimized assembly produced by GCC, it was able to
produce code competitive with higher optimization levels.
The overall structure of this superoptimizer, where program slices
are extracted, canonicalized, checked against a cache, and then
optimized in the case of a cache miss, is very similar to \tool{}, but
there are many differences in the details, particularly in \tool's
slice extractor which allows its synthesis specification to
approximate the original code's effect much more closely.
Another assembly superoptimizer, STOKE~\cite{stoke, stoke-fp,
  conditionally}, is not as closely related; it is based on randomly
perturbing assembly-language functions.
STOKE can potentially perform transformations that Minotaur cannot,
but we believe that its results are more difficult to translate into
standard peephole optimizations than are Minotaur's.

Several recent projects have focused not on optimizing individual
programs but rather on generating program rewrite rules.
OptGen~\cite{optgen} finds scalar peephole optimizations that meet
a specified syntactic form.
Even at small rewrite sizes, it was able to find numerous
optimizations that were missing from the 2015 versions of GCC and
LLVM\@.
VeGen~\cite{vegen} generates SLP vectorization rules---an SLP
vectorizer~\cite{slp} merges a set of scalar operations into vector
instructions.
VeGen parses the Intel Intrinsics Guide~\cite{intelguide} and uses this
to build pattern matchers for x86 vector instructions.
VeGen applies the pattern matchers to an input scalar program, and
replaces scalar expressions with vector instructions when it
finds a profitable match.
VeGen uses syntactic pattern matching rather than solver-based
equivalence/refinement checking.
Diospyros~\cite{diospyros} is another vector rewrite rule generator,
it takes an equality saturation~\cite{equalitysat} approach and uses a translation
validator to reject unsuitable candidates.
As an equality saturation-based tool, Diospyros builds its search space
with existing rewrite rules.

Program synthesis---generating implementations that conform to
a given specification---is intimately related to superoptimization.
Rake~\cite{rake} performs instruction selection for vectorized
Halide~\cite{halide} expressions using a two stage synthesis
algorithm.
First, Rake synthesizes a data-movement-free sketch~\cite{sketch}, and
then in the second stage it concretizes data movement for the
sketch via another synthesis query.
Rake targets Hexagon DSP processors~\cite{hexagon} which share some functionally
similar SIMD instructions with x86\@.
Cowan et al.~\cite{ml_syn} synthesized quantized machine learning
kernels.
Their work introduces two sketches: a compute sketch, which computes a matrix
multiplication, and a reduction sketch that collects the computation
result to the correct registers.
It relies on Rosette~\cite{rosette} to generate an efficient NEON~\cite{neon}
implementation that satisfies the specifications for those two
sketches.
Swizzle Inventor~\cite{swizzleinventor} is another tool built on
Rosette; it synthesizes data movement instructions for a GPU compute
kernel, and it requires user-defined sketches describing the
non-swizzle part of the program.
MACVETH~\cite{sparse} generates high-performance vector packings of
regular strided-access loops, by searching for a SIMD expression that
is equivalent to a gather specification.
All of these works show good performance results, but they focus on
relatively narrow tasks, whereas \tool{} attempts to improve SIMD
programs in general.

Most previous superoptimizers and program synthesizers use simple
cost models.
For example, Souper~\cite{souper} assigns each kind of instruction a
weight and uses the weighted sum as the cost of a rewrite.
This kind of cost model is not a very good predictor of performance
on a modern out-of-order processor.
\tool{} and MACVETH~\cite{sparse} use the LLVM-MCA~\cite{llvmmca}
microarchitectural performance analyzer, which can still lead to
mispredictions, but it is generally more accurate than simple
approaches are.

\section{Conclusion}
\label{sec:conc}

We created \tool{} because we noticed that LLVM appeared to be missing
relatively obvious optimizations in code containing both its portable
vector instructions and also its platform-specific intrinsic
functions that provide direct access to hardware-level primitives.
\tool{} cuts loop-free DAGs of instructions---including branches and
memory operations---out of LLVM functions and then attempts to
synthesize better implementations for them.
When improved code is found, the optimization is performed and also
the synthesis result is cached.
On the libYUV test suite, \tool{} gives speedups up to 1.64x,
with an average speedup of 2.2\%.

\begin{acks}
  The authors are indebted to Nuno P.\ Lopes for creating Alive2, for
  structuring it in such a way that we could programmatically access
  its functionality, and for helping us to use it effectively.
  Alexander Brauckmann and Michael O'Boyle generously provided access
  to a collection of loop kernels extracted using their tool, FlexC.
  Alastair Reid, Fabian Giesen, Sam Elliott, Raimondas Sasnauskas,
  Pavel Panchekha, Manasij Mukherjee, Tanmay Tirpankar, and anonymous
  reviewers all provided invaluable feedback on drafts of this paper.
This material is based upon work supported by the
\grantsponsor{NSF}{National Science Foundation}{https://www.nsf.gov/} under Grant
No.~\grantnum{NSF}{1955688}.
\end{acks}

\bibliographystyle{ACM-Reference-Format}
\bibliography{main}

\end{document}